# Elegant vector normal modes at a dielectric interface

**Wojciech Nasalski**

Institute of Fundamental Technological Research, Polish Academy of Sciences
Świętokrzyska 21, 00-049 Warsaw, Poland
wnasal@ippt.gov.pl

Reflection and transmission of narrow beams at a dielectric interface is analysed. It is confirmed that for arbitrary incidence two types of beams – Elegant Hermite-Gaussians of linear polarization and Elegant Laguerre-Gaussians of circular polarization, both defined in the interface plane – can be treated as vector normal modes of the interface. Excitation of higher-order modes by cross-polarization coupling is described by changes of mode indices induced by generalised transmission and reflection matrices. It is explicitly shown that, off-axis in general, spectral placements of the excited optical vortices are determined by elements of these matrices, depended in turn on beam incidence. Numerical simulations of beam reflection entirely confirm theoretical predictions. The current paper is a continuation and extension of a previous work by W. Nasalski [Phys. Rev. E **74**, 056613 (2006)].

# 1. INTRODUCTION

It is known that a field structure of a bounded beam of light reflected or refracted at a planar interface differs substantially from that of an incident beam in its intensity, phase and polarization [1, 2]. For sufficiently wide beams these differences can be described approximately by beam frame displacements from predictions of geometric optics (g-o) and by appropriate coordinate scaling (see [3] and references therein). Among these effects the known the most are the longitudinal Goos-Hänchen [4, 5] and transverse Imbert-Fedorov [6, 7] shifts. However this geometrical approach works well for three-dimensional (3D) beams of cross-section diameters not less than around ten wavelegths. Moreover, even in this range and in spite of the case of the fundamental Gaussian, only higher-order non-singular beams, the field of which can be factorised into two two-dimensional (2D) beam fields, like Hermite-Gaussian (HG) beams, can be in principle treated efficiently in this way [3].

Recently, beams of helical phase structure are under intense research, because of many interesting applications ranging from quantum communication to visualisation in biology and medicine. However Lagurre-Gaussian (LG) beams for example, even of diameter of the order of hundreds wavelegths or more, suffer at the dielectric interface from such huge distortions, that the standard geometrical displacements of their field distributions are not sufficient to describe them even approximately [8]. It seems that reasons for these distortions are inhibited, besides of beam elliptic polarization, in the very specific interrelations between beam angular momentum (AM), beam spectrum and placement of optical vortices embedded in the beam field [9]. A remedy in this situation may be to search for the information on the beam directly in its spectral, instead of direct spatial, field distribution [10, 11]. It was recently shown that inspection of the information encoded in spiral spectra of light beams has promising potential in optical imaging [11].

A full-wave vectorial analysis of beams of this type at the dielectric interface was recently reported [12]. The method was applied to HG and LG beams in their elegant version devised some time ago by Siegman [13, 14]. Such beams appeared to be good candidates for normal modes at any planar dielectric structure. It was shown that cross-polarization coupling (XPC) between orthogonal beam components is responsible for excitation of higher-order beams at such a structure, in the case of their sufficiently narrow cross-sections. Although the method presented in [12] is quite general, only the problem of beam transmission under normal incidence has been there discussed numerically in deep. This contribution extends those results, discussed here for arbitrary beam incidence in parallel in the configuration and spectral domains. Similarities and differences in behaviour of HG and LG beams at the interface will be carefully discussed, basic mechanism of optical vortices excitation will be described in detail and off-axis placements of excited optical vortices will be exactly determined in the beam spectra. Numerical examples showing reflection of the beams with their radius of the order of one wavelength, that is narrow enough to be below the paraxial limit, will be presented.

In section 2 basic relations describing elegant beams will be shortly outlined. In sections 3 and 5 expressions for transmission and reflection matrices will be given. Details of their derivation were presented in [12], which in turn follows earlier author's analyses published in another context [15-17], where the idea of the XPC beam-interface interactions was first explicitly introduced in [16]. Results of independent numerical simulations, based on direct integration of Maxwell equations at the interface – not simply on the equations derived, will be also presented in sections 4 and 6. They confirm the outcome of the analysis obtained on a purely analytical level. The interface will be considered as homogeneous, lossless and isotropic. Only results on beam internal reflection at the interface, with dielectric contrast equal two, will be given, as those on beam refraction were already reported in [12]. Finally, additional comments are given in section 7 and main conclusions summarise the paper in section 8.

## 2. BASIC DEFINITIONS FOR ELEGANT BEAMS

There are two levels of spectral analysis of beam fields at the interface. At the geometric-optical (g-o) level the standard Snell and Fresnel laws govern a single plane wave incident at a polar incidence angle $\vartheta^{(i)}$. On the second level plane waves are spectral ingredients of the 3D beams and therefore dependent, besides $\vartheta^{(i)}$, on an azimuthal incidence angle $\varphi$. In this paper the beams are treated analytically and exactly on the second level, where the well-known Fresnel coefficients should be replaced in the spectral or momentum domain by appropriately defined transmission and reflection matrices, dependent on both angles $\vartheta^{(i)}$ and $\varphi$. Subsequently, by use of the standard 2D Fourier transform, the beam fields are obtained numerically in the configuration or direct domain, with incidence angles $\theta^{(i)}$ and $\psi$ (see figure. 1).

In general, a complex electric field vector $\underline{E}_\perp^{(b)}$ transverse to the normal to the interface is obtained by the decomposition consisted of *suitably defined* mode solutions $\underline{G}_{c,d}^{(b)}$ to the problem:

$$\underline{E}_\perp^{(b)}(X,Y,Z) = \sum_{c,d} a_{c,d}^{(b)} \underline{e} \circ \underline{G}_{c,d}^{(b)}(X,Y,Z'), \qquad (1)$$

where $a_{c,d}^{(b)}$ are expansion coefficients, $c$ and $d$ are expansion indices, $Z' = Z - Z_w$, as the beam waists are assumed be placed in general in planes $Z = Z_w$ parallel to the interface, $b = i,t,r$ indicate the incident, transmitted (refracted) and reflected beam, respectively (see figure 1). The polarization 2D versor $\underline{e}$ is defined in the interface plane $X - Y$. For uniform beam polarization a scalar wave approach suffices to describe the beam field in its paraxial region with polarization components of $\underline{G}_{c,d}^{(b)}$ regarded as modes of the optical system. Still the true modes at the interface are vectors not scalars because the interface couples the polarization and spatial characteristics of beam fields [12, 15-17].

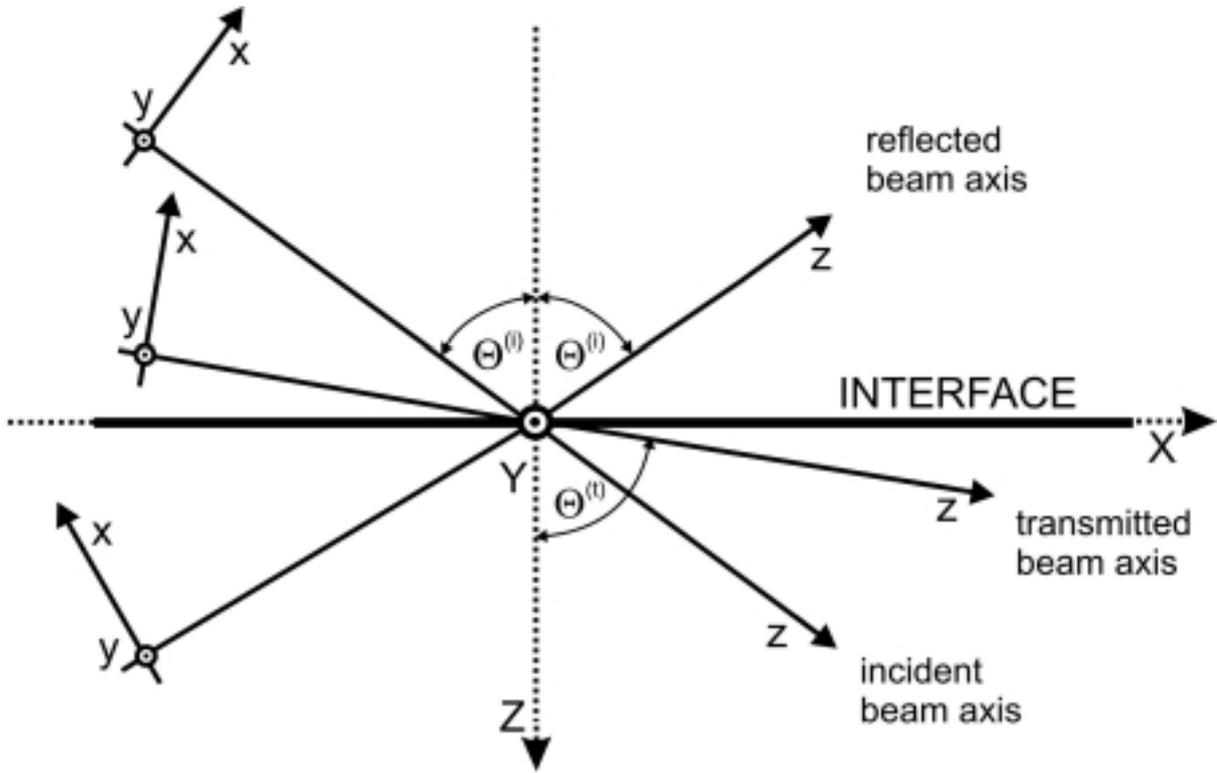

**Figure 1.** Interface $OXYZ$ and beam $Oxyz$ reference frames viewed in a main incidence plane $X - Z$; local planes of incidence are given by rotation of the plane $X - Z$ by an azimuthal angle $\psi$ around the axis $Z$. Beam waist centres are placed in centres of beam frames, incidence of internal reflection is assumed.

Only monochromatic fields are considered with the propagation term $\exp(-i\omega t + \mathrm{i} k^{(b)} Z'/\cos\theta^{(b)})$ assumed and suppressed in all field expressions, where and $k^{(b)}$ are wave numbers of the respective beam fields. Therefore, in the paraxial regions of beams, the components of $\underline{E}_\perp^{(b)}$ and $\underline{G}_{c,d}^{(b)}$ are slowly varying vector field envelopes (SVE) and the Z component is fixed by the transversality constraint on the beam field.

Our aim in this paper is to show and discuss examples of these *suitably defined* partial solutions to the problem. They satisfy paraxial approximation to Maxwell equations in two half-spaces $Z > 0$ and $Z < 0$, together with the continuity relations of transverse field components at the interface assumed as placed at the plane $Z = 0$. If the modes constitute a complete and orthogonal or bi-orthogonal set of square integrable functions, they can be called as *normal modes* of the interface. Therefore for any normal mode $\underline{G}_{c,d}^{(b)}$ the expansion (1) resolves into only one term representing field $\underline{E}_\perp^{(b)}$ of all three beams $b = i, t, r$.

Still, for uniform (transverse) polarization, we can consider scalar modes for free space propagation. Two sets of such modes or beams will be considered further – Elegant Hermite-Gaussian (EHG) beams and Elegant Laguerre-Gaussian (ELG) beams [13, 14, 18, 19]. They specify solutions to the problem in the two coordinate systems: Cartezian $OXYZ$ of rectangular symmetry and cylindrical $Or_\perp \psi Z$ of cylindrical symmetry. It appears that the elegant beams are particularly suitable in description of beam fields at the interface provided that they are defined at the interface plane $X - Y$ instead, as usual, in the beam transverse plane $x - y$ (see figure 1) [12]. It will be shown that the fields of the EHG and ELG beams fit very well to transmission and reflection matrices defined in the interface plane.

Let us consider normal incidence ($\theta^{(i)} = 0$) first. In this case the two planes $X - Y$ and $x - y$ coincide. In the Cartezian coordinates the Fock (paraxial) wave equation reads

$$\{ik^{(b)}\partial_Z + \tfrac{1}{2}(\partial_X^2 + \partial_Y^2)\} G_{m,n}^{(EH)}(X,Y,Z') = 0, \qquad (2)$$

with the Elegant solution in the form of the EHG beam mode $G_{m,n}^{(EH)}$. The solution is specified by the $X$ and $Y$ indices given by two integers $m$ and $n$, respectively, their order $N^{(EH)} = m + n$ and the beam waist placement in the plane $Z = Z_w$ [12, 19]:

$$G_{m,n}^{(EH)}(X,Y,Z') = (w_w)^{m+n} \partial_X^m \partial_Y^n G_{0,0}^{(EH)}(X,Y,Z'). \qquad (3)$$

$$G_{0,0}^{(EH)}(X,Y,Z') = (w_w/v(Z'))^2 \exp\bigl(-(1/2)(X^2 + Y^2)/v^2(Z')\bigr), \qquad (4)$$

$$v^2(Z') = w_w^2(1 + iZ'/z_D), \qquad (5)$$

where $z_D = kw_w^2$ is the diffraction length or Rayleigh range, $w_w$ and $v(Z')$ are the waist and complex radii of the fundamental Gaussian mode $G_{0,0}^{(EH)}$, respectively. The Gaussian mode is assumed in this work as symmetric with circular cross-section, i.e. with the same waist radii $w_{wX} = w_w = w_{wY}$ in X and Y directions in each plane parallel to the interface plane. For elliptic cross-section of the beam, i.e. with different beam radii $w_{wX}$ and $w_{wY}$ modification in all expressions of this paper is straightforward.

The EHG functions $G_{m,n}^{(EH)}$ are expressed by the Hermite polynomials $H_m(x)$ of order m:

$$G_{m,n}^{(EH)}(X,Y,Z') = (-w_w)^{m+n} H_m(2^{-1/2} X/v) H_n(2^{-1/2} Y/v) G_{0,0}^{(EH)}(X,Y,Z'), \qquad (6)$$

$$(-1)^m H_m(x) \exp(-x^2) = (d^m/dx^m) \exp(-x^2). \qquad (7)$$

The amplitude of the EHG beam is determined by the condition $G(0,0,0) = 1$. The form of this condition is introduced here to make the analytical results as concise as possible. However, any beam field normalisation is also possible to be introduced into this analysis.

Similar definitions can be applied in the complex coordinates specific to systems of cylindrical symmetry: $\varsigma = 2^{-1/2}(X + iY)$ and its complex conjugate $\bar{\varsigma}$. Then the paraxial wave equation reads [12, 19]:

$$\{ik^{(b)}\partial_Z + \partial_\varsigma \partial_{\bar{\varsigma}}\} G_{p,l}^{(EL)}(\varsigma,\bar{\varsigma},Z') = 0, \tag{8}$$

with Elegant solution given in a form of ELG modes:

$$G_{p,l}^{(EL)}(\varsigma,\bar{\varsigma},Z') = w_w^{2p+l} \partial_\varsigma^p \partial_{\bar{\varsigma}}^{p+l} G_{0,0}^{(EL)}(\varsigma,\bar{\varsigma},Z'), \tag{9}$$

$$G_{0,0}^{(EL)}(\varsigma,\bar{\varsigma},Z') = (w_w/v(Z'))^2 \exp(-\varsigma\bar{\varsigma}v^{-2}(Z')), \tag{10}$$

of the radial $p$ and azimuthal $l$ indices contributing to the beam order $N^{(EL)} = 2p + l$. Note that $G_{0,0}^{(EL)} = G_{0,0}^{(EH)}$ for $X = 2^{-1/2}(\varsigma + \bar{\varsigma})$ and $Y = -i2^{-1/2}(\varsigma - \bar{\varsigma})$ and the ELG modes are expressed by the associated Laguerre polynomials $L_p^l(x)$:

$$\begin{aligned} G_{p,l}^{(EL)}(\varsigma,\bar{\varsigma},Z') = &(-1)^{p+l}(\varsigma_\perp/v)^l (v/w_w)^{-(2p+l)} \\ &\times p! L_p^l(\varsigma_\perp^2/v^2) G_{0,0}^{(EL)}(\varsigma,\bar{\varsigma},Z') \exp(il\psi) \end{aligned}, \tag{11}$$

$$p! L_p^l(x)[x^l \exp(-x)] = [(d/dx)x]^p [x^l \exp(-x)], \tag{12}$$

where $\varsigma_\perp = (\varsigma\bar{\varsigma})^{1/2}$ and $\psi = (1/2)\ln(\varsigma/\bar{\varsigma})$. The azimuthal index or winding number $l$ is any integer number associated with the Z component of the orbital angular momentum (OAM) carried by the ELG beam. When $l \neq 0$ it is equal to a topological charge of a screw phase dislocation or an optical vortex, which the beam field contains. The radial index or node number $p$ is a non-negative integer, which determines a radially symmetric beam structure with, besides the fundamental Gaussian case, zero on-axis intensity.

For oblique incidence ($\theta^{(i)} \neq 0$) the above definitions of the Elegant beams still remain valid provided that the projection of the transverse (to the propagation direction) field distribution of the fundamental Gaussian on the interface, or in general on the plane $Z = Z_w$ parallel to the interface, was performed first [12]. In this plane different plane waves acquire additional phase shifts dependent of the incidence angles $\vartheta^{(i)}$ and $\varphi$. These changes modify the commonly understood definitions of the Elegant beams. Therefore such the beams should be called rather as the *Projected Elegant - PEHG and PELG – beams* in the context of their oblique incidence. Note that for oblique incidence the modes are still considered here as symmetric in the beam transverse plane, with equal waist radii $w_{wx} = w_w = w_{wy}$ in planes transverse to the beam propagation direction and non-equal radii $w_{wX} = w_w/\cos\theta^{(i)}$ and $w_{wY} = w_w$ in the interface. However, generalisation of this analysis to asymmetric cases is straightforward.

Although paraxial beam fields are considered in this section, all further analytical calculations will be exact as carried out in a single plane – the interface plane. Therefore one can, starting from the beam field relations obtained in this plane, rebuild exact non-paraxial beam fields in the vicinity of this plane ruled exactly by a full set of Maxwell equations. For this reason a small cross-sectional diameter of beams, corresponding to $kw_w = 2\pi$ and being below the paraxial limit, will be taken in all numerical simulations.

## 3. BEAM-INTERFACE RELATIONS IN RECTANGULAR COORDINATES

Let us start from the Fresnel transmission and reflection coefficients $t_p$, $r_p$ and $t_s$, $r_s$ for a single plane wave of an arbitrary polarization parameter $\tilde{\chi}^{(i)}_{(p,s)} = \tilde{E}^{(i)}_p / \tilde{E}^{(i)}_s$ specified by its p and s polarization components $\tilde{E}^{(i)}_p$ and $\tilde{E}^{(i)}_s$, respectively. The plane wave is incident on the interface under the incidence angle $\vartheta^{(i)}$. In the linear TM/TE polarization basis $\underline{e}_{(X,Y)} = [\underline{e}_X, \underline{e}_Y]$ defined in the interface plane $X-Y$ these coefficients form diagonal elements of the transmission and reflection matrices

$$\underline{\underline{t}}_{(p,s)} = \begin{bmatrix} \eta t_p & 0 \\ 0 & t_s \end{bmatrix}, \tag{13}$$

$$\underline{\underline{r}}_{(p,s)} = \begin{bmatrix} r_p & 0 \\ 0 & r_s \end{bmatrix}, \tag{14}$$

where $\eta = \cos\vartheta^{(t)}/\cos\vartheta^{(i)}$. It is stipulated that $r_p = r_s = 1$ for critical incidence of total internal reflection (TIR). Then, in the interface, the continuity relations of tangent field components are given by:

$$\underline{\underline{t}}_{(p,s)} = \underline{\underline{1}} + \underline{\underline{\sigma}}\, \underline{\underline{r}}_{(p,s)}, \tag{15}$$

where $\underline{\underline{1}}$ is a unit matrix and $\underline{\underline{\sigma}}$ stand for diagonal matrix of $\sigma_{XX} = -1 = -\sigma_{YY}$. Note that matrices (13)-(14) do not depend of the azimuthal angle $\varphi$ and are defined in one, usually taken as a main, incidence plane [15].

However, 3D beams are composed of infinite number of plane waves and each plane wave is distinguished by two angles of incidence - a polar angle $\vartheta$ and an azimuthal angle $\varphi$, where the latter defines a local incidence plane attributed to this plane wave [15]. The matrices $\underline{\underline{t}}_{(p,s)}$ and $\underline{\underline{r}}_{(p,s)}$ are then replaced by the matrices $\underline{\underline{t}}_{(X,Y)}$ and $\underline{\underline{r}}_{(-X,Y)}$, still diagonal but this time dependent on both incidence angles $\vartheta^{(i)}$ and $\varphi$ [12, 17]:

$$\underline{\underline{t}}_{(X,Y)} = \begin{bmatrix} \eta t_{TM} & 0 \\ 0 & t_{TE} \end{bmatrix} = \begin{bmatrix} \eta t_p + \Delta_{TM} & 0 \\ 0 & t_s + \Delta_{TE} \end{bmatrix}, \quad (16)$$

$$\underline{\underline{r}}_{(-X,Y)} = \begin{bmatrix} r_{TM} & 0 \\ 0 & r_{TE} \end{bmatrix} = \begin{bmatrix} r_p - \Delta_{TM} & 0 \\ 0 & r_s + \Delta_{TE} \end{bmatrix}, \quad (17)$$

with their elements expressed by the Fresnel coefficients $\eta t_p$, $t_s$, $r_p$, $r_s$ modified by

$$\Delta_{TM} = 2 t_{CX} k_\perp^{-2} k_Y [\widetilde{\chi}_{(X,Y)}^{(i)}{}^{-1} k_X - k_Y], \quad (18)$$

$$\Delta_{TE} = 2 t_{CX} k_\perp^{-2} k_Y [\widetilde{\chi}_{(X,Y)}^{(i)}{}^{+1} k_X + k_Y], \quad (19)$$

where $\underline{k} = [k_\perp, k_Z]^T = [k_X, k_Y, k_Z]^T$ and the XPC coefficients

$$t_{CX} = \tfrac{1}{2}(\eta t_p - t_s), \quad (20)$$

$$r_{CX} = \tfrac{1}{2}(r_p + r_s), \quad (21)$$

$r_{CX} = -t_{CX}$. Their role in the XPC interactions will appear evident further.

The matrices $\underline{\underline{t}}_{(X,Y)}$ and $\underline{\underline{r}}_{(-X,Y)}$ could be given in their diagonal form due to introduction of the complex polarization parameter:

$$\widetilde{\chi}_{(X,Y)}^{(i)} = \widetilde{E}_X^{(i)} / \widetilde{E}_Y^{(i)}, \quad (22)$$

dependent in general on $\vartheta^{(i)}$ and $\varphi$, and equal to $\infty$, 0, -i and i for TM, TE, CR and CL beam polarization states of the incident beam, respectively. These matrices relate spectral components $\underline{\widetilde{E}}_{(X,Y)}^{(b)} = [\widetilde{E}_X^{(b)}, \widetilde{E}_Y^{(b)}]^T$, b = i, t and $\underline{\widetilde{E}}_{(-X,Y)}^{(r)} = [-\widetilde{E}_X^{(r)}, \widetilde{E}_Y^{(r)}]^T$ of the beam fields at the interface:

$$\underline{\widetilde{E}}_{(X,Y)}^{(t)} = \underline{\underline{t}}_{(X,Y)} \underline{\widetilde{E}}_{(X,Y)}^{(i)}, \quad (23)$$

$$\underline{\widetilde{E}}_{(-X,Y)}^{(r)} = \underline{\underline{r}}_{(-X,Y)} \underline{\widetilde{E}}_{(X,Y)}^{(i)} \quad (24)$$

and fulfil there the field continuity relations:

$$\underline{\underline{t}}_{(X,Y)} = \underline{\underline{1}} + \underline{\underline{\sigma}}\, \underline{\underline{r}}_{(-X,Y)}. \quad (25)$$

Therefore, for incidence of 3D beams, the Fresnel matrices $\underline{\underline{t}}_{(p,s)}$ and $\underline{\underline{r}}_{(p,s)}$ are replaced by their 3D generalisations $\underline{\underline{t}}_{(X,Y)}$ and $\underline{\underline{r}}_{(-X,Y)}$ and the Fresnel coefficients $\eta t_p$, $t_s$, $r_p$, $r_s$ are replaced by their 3D generalisations $\eta t_{TM}$, $t_{TE}$, $r_{TM}$, $r_{TE}$. Their analytical form is valid for any plane wave of which the incident beam is composed. They are interrelated through the continuity relations at the interface: Eq. (15) given in the main plane of incidence $\varphi=0$ and Eq. (25) given in any the local plane of incidence defined by arbitrary value of $\varphi$.

It is convenient to relate the matrices $\underline{\underline{t}}_{(X,Y)}$ and $\underline{\underline{r}}_{(-X,Y)}$ directly to the azimuthal angle $\varphi = \arctan(k_Y/k_X)$ through the relations $\cos 2\varphi = (k_X^2 - k_Y^2)k_\perp^{-2}$ and $\sin 2\varphi = 2k_X k_Y k_\perp^{-2}$. Then they are separated into following diagonal/antidiagonal components [17]:

$$\underline{\underline{t}}_{(X,Y)} = \underline{\underline{t}}_{(p,s)} + t_{CX}\begin{bmatrix} 0 & 1 \\ 1 & 0 \end{bmatrix}\sin 2\varphi + 2t_{CX}\begin{bmatrix} -1 & 0 \\ 0 & 1 \end{bmatrix}\sin^2\varphi, \qquad (26)$$

$$\underline{\underline{r}}_{(-X,Y)} = \underline{\underline{r}}_{(p,s)} + r_{CX}\begin{bmatrix} 0 & 1 \\ -1 & 0 \end{bmatrix}\sin 2\varphi - 2r_{CX}\begin{bmatrix} 1 & 0 \\ 0 & 1 \end{bmatrix}\sin^2\varphi. \qquad (27)$$

Although the decomposition (26)-(27) is exact, the subsequent terms in this decomposition can be interpreted as the zero-order (Fresnel), first-order and second-order (with respect to $k_Y$ or $\varphi$) contributions to beam transmission and reflection. Only the first-order terms are created by the XPC effect at the interface and only these terms are dependent – through $\tilde{\chi}^{(i)}_{(X,Y)}$ - on a polarization state of the incident beam field. Certainly, besides the azimuthal angle $\varphi$ the matrices $\underline{\underline{t}}_{(X,Y)}$ and $\underline{\underline{r}}_{(X,Y)}$ depend also, through the Fresnel coefficients, on the polar angle $\vartheta^{(i)}$.

Exactly in the incidence plane ($\varphi = 0$) the first-order and second-order terms in Eqs (26)-(27) disappear and only the zero-order term, determined by rules of geometrical optics, remains. In this plane the generalised matrices (26)-(27) resolve plainly into the Fresnel or g-o matrices (13)-(14). Elements of these matrices can be further expressed by $t_C \pm t_{CX}$ and $r_C \pm r_{CX}$, where $t_C$ and $r_C$ are defined later by Eqs. (42)-(43). Still, the Fresnel matrices are diagonal. It means that in the Cartesian coordinates the g-o approach to the problem disregards the XPC effects in the beam-interface interactions.

For reasons which will be evident soon, let us assume for a while that $\varphi$ is small and approximate additionally the decomposition (26)-(27), in the vicinity of the incidence plane and up to the second-order terms, by expansion of the beam field with respect to $\vartheta^{(i)}$ around $\theta^{(i)}$. Therefore, for $\varphi \cong 0$ and $\vartheta^{(i)} - \theta^{(i)} \cong 0$, both expansions, longitudinal with respect to $\vartheta^{(i)}$ and transverse with respect to $\varphi$, lead to the well known description of beams in terms of the first-order and the second-order effects of nonspecular transmission and reflection: longitudinal - along the $X$ axis and transverse – along the $Y$ axis [3, 15-17]. It was numerically verified that this beam description is valid as well for the transmitted near-beam-field at TIR [20] or even in cases of nonlinear interfaces [21].

Meanwhile the longitudinal effects depend on derivatives with respect to $k_X$ of the Fresnel coefficients coefficients $\eta t_p$, $t_s$, $r_p$ and $r_s$, the transverse effects are proportional to derivatives of $\Delta_{TM}$ and $\Delta_{TE}$ with respect to $k_Y$ [15-17]. All the first-order and the second-order effects of nonspecular reflection are evaluated with respect to predictions of geometrical optics. The approximation $\varphi \cong 0$ and $\vartheta - \theta_i \cong 0$ provides the clear geometrical description of beam deformations of beam centre displacements, beam frame rotations and beam coordinate scaling (see Eqs (12)-(19) and (A.1)-(A.4) in Ref. [16] or Eqs (18)-(22) in Ref. [17]). Moreover, these nonspecular effects can be further separated into changes of the beam complex amplitude and polarization (see Eqs (30)-(37) in Ref. [17]).

Under oblique incidence all the first-order and second-order beam shifts produce composite beam displacements in the plane of interface independently in the longitudinal and transverse directions. Due to the angular shift of the beam direction these composite displacements directly depend on the propagation distance (from its actual waist position) of the incident beam and are enhanced when the beam waist centre is placed far from the interface (cf. Eq. (15) in Ref. [16]). The role of beam propagation enhancement of these displacements has been also recently confirmed in the context of their measurements by weak measurement techniques [22]. Note however that for normal incidence the first-order beam displacements disappear completely [12, 16].

## 4. HERMITE-GAUSSIAN MODES OF THE INTERFACE

Eqs (26)-(27) are valid for arbitrary plane wave of arbitrary polarization. However, the transmission and reflection matrices show distinct symmetry with respect to the incidence ($Y = 0$) and transverse ($X = 0$) planes and their action is particularly specific in the case of incident beams possessing the same type of symmetry with respect to their shape and polarization. Within this class of beams the EHG/PEHG beams of TM or TE linear polarization are of particular importance.

The EHG/PEHG beams are defined in the configuration domain by differentiation of the fundamental Gaussian beam field (see Eq. 3). In the spectral domain these definitions are equivalent to the algebraic expression:

$$\widetilde{G}_{m,n}^{(EH)}(k_X, k_Y, Z') = (iw_w)^{m+n} k_X^m k_Y^n \widetilde{G}_{0,0}^{(EH)}(k_X, k_Y, Z'), \tag{28}$$

$$\widetilde{G}_{0,0}^{(EH)}(k_X, k_Y, Z') = 2\pi \exp\left[-(1/2)(k_X^2 + k_Y^2)v^2(Z')\right]. \tag{29}$$

As it was stipulated in section 2 the beam amplitude is set by the condition $\widetilde{G}(0,0,0) = 2\pi$.

Let us first consider normal incidence of the beam $\underline{\widetilde{E}}^{(i)} = [\widetilde{a}_X, \widetilde{a}_Y]^T \widetilde{G}_{m,n}^{(EH)}$ of the EHG mode shape and of arbitrary polarization $\widetilde{\chi}^{(i)}(X,Y) = \widetilde{a}_X / \widetilde{a}_Y$ specified by the beam components $\widetilde{a}_X$ and $\widetilde{a}_Y$, in general complex and dependent on $k_X$ and $k_Y$ in the case of nonuniform polarization. We assume that the impact of the second-order term in the decomposition (26)-(27) is negligibly small, as previous numerical simulations confirm this assumption [12]. Then, within this first-order approximation, Eqs (26)-(27) yield the transmitted and reflected beam fields of the form [12]:

$$\begin{bmatrix} \widetilde{E}_X^{(t)} \\ \widetilde{E}_Y^{(t)} \end{bmatrix} \cong \begin{bmatrix} \eta t_p \widetilde{a}_X \\ t_s \widetilde{a}_Y \end{bmatrix} \widetilde{G}_{m,n}^{(EH)} - 2t_{CX}(k_\perp w_w)^{-2} \begin{bmatrix} \widetilde{a}_Y \\ \widetilde{a}_X \end{bmatrix} \widetilde{G}_{m+1,n+1}^{(EH)}, \tag{30}$$

$$\begin{bmatrix} -\widetilde{E}_X^{(r)} \\ \widetilde{E}_Y^{(r)} \end{bmatrix} \cong \begin{bmatrix} r_p \widetilde{a}_X \\ r_s \widetilde{a}_Y \end{bmatrix} \widetilde{G}_{m,n}^{(EH)} - 2r_{CX}(k_\perp w_w)^{-2} \begin{bmatrix} \widetilde{a}_Y \\ -\widetilde{a}_X \end{bmatrix} \widetilde{G}_{m+1,n+1}^{(EH)}. \tag{31}$$

Both indices in the beam component of the same polarization as that of the incident beam remain unchanged ($m \to m$ and $n \to n$). However the beam component of the opposite (orthogonal) polarization shows clearly the action of the XPC effect [12, 16]. Both indices in this component are increased by one ($m \to m+1$ and $n \to n+1$) and thus the incident beam order $N^{(HG)} = m+n$ is increased there by two ($N^{(HG)} \to N^{(HG)} + 2$). Therefore, for incidence of EHG beams of a *single* TM or TE polarization component, the reflected beams are composed also of the EHG beams in the *two* TM and TE polarization components, with their indices exactly specified. The relations (30)-(31) are derived in the spectral domain. The beam field distribution in the direct domain is obtained by the standard 2D Fourier transform of these relations.

In figure 2 the case of normally incident narrow ($kw_w = 2\pi$) EHG$_{3,3}$ beam of TE polarization, collimated at the interface ($z = 0 = Z_w$), is shown in parallel in the configuration and spectral domains. Numerical simulations entirely confirm the predictions given by Eqs (30)-(31) – meanwhile the reflected beam TE component mimics the incident TE beam field intensity in a form of the mode EHG$_{3,3}$, the opposite TM component acquires additional shift by one in both indices of the EHG beam mode, what yields the EHG$_{4,4}$ mode structure. For the narrow incident beam its spectrum is sufficiently wide to make the XPC effect strong.

In spite of the fact that the XPC reflection coefficient $r_{CX}$ equals exactly zero for normal incidence $\vartheta = 0$, the beam TM component is only one order in beam intensity weaker than the reflected beam TE component. As usual for normal incidence, no beam shifts as well as no significant beam deformations are observed. This clear beam intensity structure is typical for beam-interface configurations symmetric with respect to the incident $Y = 0$ and transverse $X = 0$ planes. Results of numerical simulations shown in figure 2 for beam reflection are entirely consistent with those given for beam transmission in [12].

The symmetry inherent to normal incidence of the beams is partially broken for oblique incidence – in this case only mirror symmetry with respect to the incidence plane $Y = 0$ survives. The previous symmetry with respect to the transverse plane $X = 0$ is blurred mainly for three reasons: (i) the beam field projection on the interface plane $Z = 0$, (ii) the asymmetry of the Fresnel coefficients with respect to $\vartheta^{(i)}$ and (iii) the narrow azimuthal range of the incident beam spectrum. The asymmetry of the Fresnel coefficients is particularly enhanced close to critical incidence, where these coefficients become even singular. Just this case is shown in figure 3 for reflection of the PEHG$_{33}$ beam. Evidently, the beam spectrum shown in

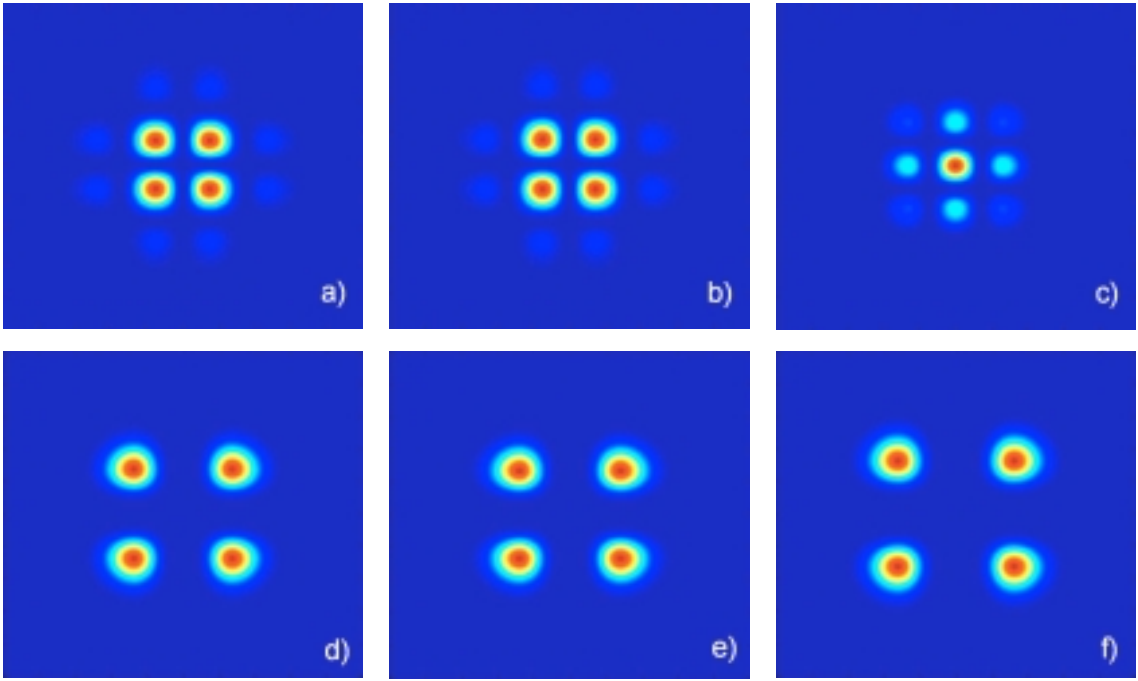

**Figure 2.** Intensity transverse distribution of the EHG beam components evaluated in the configuration (a)-(c) and spectral (d)-(f) domains in the plane of the interface. The case of normal incidence ($\theta^{(i)} = 0$) of the beam collimated at the interface is displayed for: (a) the incident beam of the $EHG_{3,3}$ pattern and of TE polarization, (b) the reflected beam TE component of the $EHG_{3,3}$ pattern, (c) the reflected beam TM component of the $EHG_{4,4}$ pattern. Counterparts of the figures (a)-(c) in the spectral domain are depicted in the figures (d)-(f), respectively. The EHG patterns in the figures (b), (e) and (c), (f) are modified by the reflection coefficients $r_s$ and $r_{CX}$, respectively.

figures 3d-3f are modified, rather slightly by the beam projection and rather strongly by the discontinuity in derivatives of the Fresnel coefficients at the critical incidence. Below critical incidence the spectrum is weak, above critical incidence the spectrum is strong and in addition shifted in phase. Still, however, the main mode structure typical for normal incidence survives in the transverse *Y* direction.

The reflected beam field shows different, with respect to the case of normal incidence, behaviour along the *X* direction. Because of the narrow range of the azimuthal angle $\varphi$ this case is more directly described by applying the next step: $\sin 2\varphi \cong 2\sin\varphi = 2k_Y/k_\perp$ in the first-order approximation $\sin^2\varphi \cong 0$ to the exact expressions (30)-(31) of the transmission and reflection matrices. By this additional approximation the decomposition (30)-(31) yields for the PEHG beams:

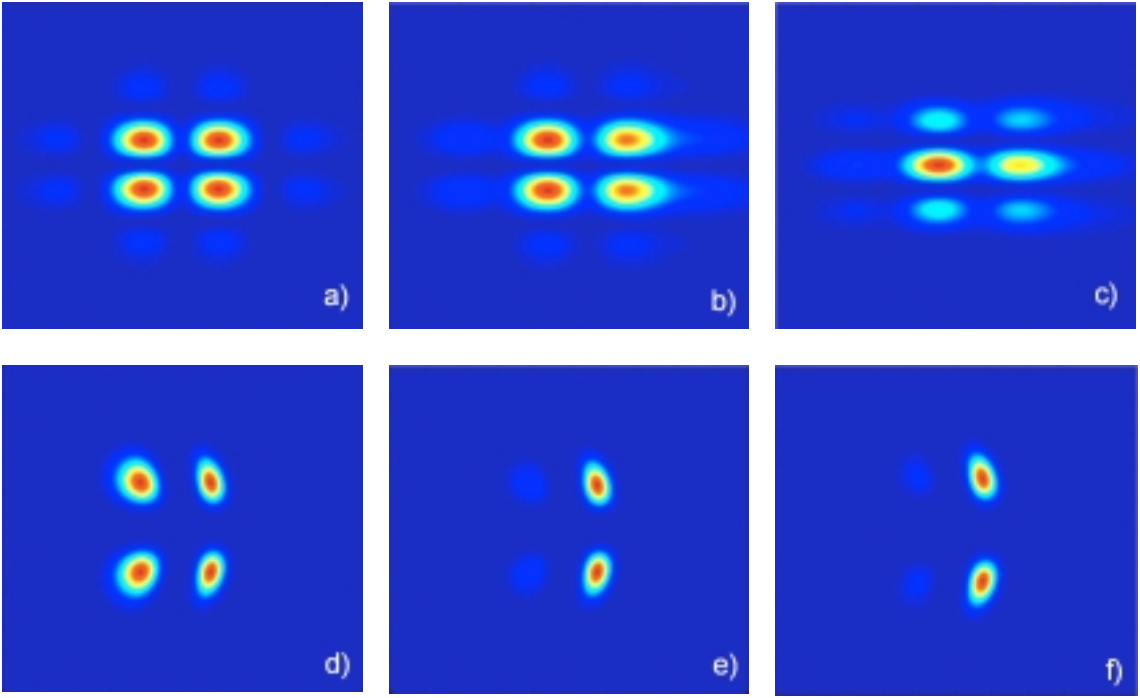

**Figure 3.** Intensity transverse distribution of the EHG beam components evaluated in the cofiguration (a)-(c) and spectral (d)-(f) domains in the interface plane - the case of critical incidence ($\theta^{(i)} = 45^o$). Besides this value all other data are the same as in figure 2.

$$\begin{bmatrix} \widetilde{E}_X^{(t)} \\ \widetilde{E}_Y^{(t)} \end{bmatrix} \cong \begin{bmatrix} \eta t_p a_X \\ t_s a_Y \end{bmatrix} \widetilde{G}_{m,n}^{(EH)} - 2it_{CX}(k_\perp w_w)^{-1} \begin{bmatrix} a_Y \\ a_X \end{bmatrix} \widetilde{G}_{m,n+1}^{(EH)}, \qquad (32)$$

$$\begin{bmatrix} -\widetilde{E}_X^{(r)} \\ \widetilde{E}_Y^{(r)} \end{bmatrix} \cong \begin{bmatrix} r_p a_X \\ r_s a_Y \end{bmatrix} \widetilde{G}_{m,n}^{(EH)} - 2ir_{CX}(k_\perp w_w)^{-1} \begin{bmatrix} a_Y \\ -a_X \end{bmatrix} \widetilde{G}_{m,n+1}^{(EH)}. \qquad (33)$$

The PEHG beams under oblique incidence behave differently than the EHG beams under normal incidence. The beam modes are still of the PEHG$_{3,3}$ shape for the TE component of the reflected beam (figure 3e). However for the opposite field component only *Y* index is increased by one and the beam mode is of the PEHG$_{3,4}$ shape. In this beam component the mode indices $(m,n)$ show transformation ($m \to m$ and $n \to n+1$) and the incident beam order $N^{(HG)} = m + n$ is increased only by one ($N^{(HG)} \to N^{(HG)} + 1$). This PEHG beam behaviour is precisely confirmed in the configuration and spectral domains by figures 3c-3f. As both Fresnel coefficients and the XPC coefficient $-r_{CX}$ as well, equal one exactly at the critical angle, the XPC beam (TM) component is only three times smaller in intensity than the g-o beam (TE) component. Therefore the creation of higher-order modes by XPC effect is much more efficient at critical incidence than at normal incidence.

In addition, the asymmetry of the Fresnel reflection coefficients $r_p$, $r_s$ with respect to $\vartheta$ leads to the reflected beam shifts in $X$ direction and their singularities at the critical incidence cause further enhancements of these shifts [16, 17]. These longitudinal shifts are clearly visible in figures 3b-3c. Moreover, the dependence of the modification terms $\Delta_{TM}$ and $\Delta_{TE}$ on the azimuthal angle $\varphi$ produces further small transverse shifts in the plane $X = 0$ [16, 17], too small, however, to be directly shown here. Still, the results of numerical simulations shown in figure 3 fully confirm analytical predictions of these shifts given in Refs [16] and [17], together with numerical simulations given there on the grounds of these equations.

Although the field decompositions (30)-(31) and (32)-(33) are only of the first-order, they appear very accurate even for the beam waist radius of the order of one wavelength. The EHG/PEHG beams of TM or TE polarization may be regarded as normal modes of the dielectric interface, as well as of any planar stratified dielectric structure. Transmission or reflection of any other beam at such the structure can be evaluated by the beam field decomposition (1) into the series of these modes in the spectral domain. In the direct domain these modes are in general modified due to dependence of the expansion coefficients on the wave vector. Still, for sufficiently wide beams, these coefficients can be treated as approximately constant and then the PEHG beams can be regarded as normal modes also in the configuration domain.

## 5. BEAM-INTERFACE RELATIONS IN CYLINDRICAL COORDINATES

The EHG/PEHG beams are normal modes at the interface only within the first-order approximation to the beam field with respect to the azimuthal angle $\varphi$. The question therefore arises whether there are other beams with different shape and polarization which can considered as normal modes and which determine the beam field *exactly*. The answer is positive - the beams symmetric with respect to cylindrical coordinates of circular polarization are such good candidates in this case [12].

Therefore let us now rewrite the expressions given previously for the EHG/PEHG beams through the straightforward unitary transformation in the circular polarization basis [12]:

$$\underline{e}_{(R,L)} = 2^{-1/2} [\underline{e}_X + i\underline{e}_Y, \underline{e}_X - i\underline{e}_Y], \tag{34}$$

and with the new polarization parameter given in this basis:

$$\widetilde{\chi}_{(R,L)}^{(i)} = \widetilde{E}_R^{(i)} / \widetilde{E}_L^{(i)} = (\widetilde{\chi}_{(X,Y)}^{(i)} - i)/(\widetilde{\chi}_{(X,Y)}^{(i)} + i). \tag{35}$$

The parameter $\widetilde{\chi}_{(R,L)}^{(i)}$ equals to $\infty$, 0, 1 and $-1$ for CR, CL, TM and TE polarization states of the incident beam, respectively. Then, the beam field vectors:

$$\underline{\widetilde{E}}_{(R,L)}^{(b)} = [\widetilde{E}_R^{(b)}, \widetilde{E}_L^{(b)}]^T = 2^{-1/2} [\widetilde{E}_X^{(b)} - i\widetilde{E}_Y^{(b)}, \widetilde{E}_X^{(b)} + i\widetilde{E}_Y^{(b)}]^T, \tag{36}$$

$$\underline{\widetilde{E}}_{(L,R)}^{(r)} = [\widetilde{E}_L^{(r)}, \widetilde{E}_R^{(r)}]^T = 2^{-1/2} [-\widetilde{E}_X^{(r)} + i\widetilde{E}_Y^{(r)}, -\widetilde{E}_X^{(r)} - i\widetilde{E}_Y^{(r)}]^T, \tag{37}$$

$b = i, t$, are interrelated at the interface by new transmission $\underline{\underline{t}}_{(R,L)}$ and reflection $\underline{\underline{r}}_{(L,R)}$ matrices:

$$\underline{\widetilde{E}}^{(t)}_{(R,L)} = \underline{\underline{t}}_{(R,L)} \underline{\widetilde{E}}^{(i)}_{(R,L)}, \tag{38}$$

$$\underline{\widetilde{E}}^{(r)}_{(L,R)} = \underline{\underline{r}}_{(L,R)} \underline{\widetilde{E}}^{(i)}_{(R,L)}, \tag{39}$$

$$\underline{\underline{t}}_{(R,L)} = \begin{bmatrix} t_{CR} & 0 \\ 0 & t_{CL} \end{bmatrix} = \begin{bmatrix} t_C + \Delta_{CR} & 0 \\ 0 & t_C + \Delta_{CL} \end{bmatrix}, \tag{40}$$

$$\underline{\underline{r}}_{(L,R)} = \begin{bmatrix} r_{CR} & 0 \\ 0 & r_{CL} \end{bmatrix} = \begin{bmatrix} r_C - \Delta_{CR} & 0 \\ 0 & r_C - \Delta_{CL} \end{bmatrix}. \tag{41}$$

The g-o coefficients take the form of averages of the Fresnel coefficients

$$t_C = \tfrac{1}{2}(\eta t_p + t_s), \tag{42}$$

$$r_C = \tfrac{1}{2}(r_p - r_s), \tag{43}$$

$t_C = 1 - r_C$, in addition modified by the XPC terms:

$$\Delta_{CR} = t_{CX} \widetilde{\chi}^{(i)\ -1}_{(R,L)} \exp(-2i\varphi), \tag{44}$$

$$\Delta_{CL} = t_{CX} \widetilde{\chi}^{(i)\ +1}_{(R,L)} \exp(+2i\varphi). \tag{45}$$

Note that $t_{CX}$ and $r_{CX}$ are the XPC transmission and reflection coefficients, respectively (see Eqs (20)-(21)).

The g-o coefficients $t_C$ and $r_C$ yield the diagonal $2 \times 2$ matrices $\underline{\underline{t}}_C = t_C \underline{\underline{1}}$ and $\underline{\underline{r}}_C = r_C \underline{\underline{1}}$ and the field continuity relations at the interface now read

$$\underline{\underline{t}}_C = \underline{\underline{1}} - \underline{\underline{r}}_C \tag{46}$$

$$\underline{\underline{t}}_{(R,L)} = \underline{\underline{1}} - \underline{\underline{r}}_{(L,R)} \tag{47}$$

in the incidence plane $\varphi = 0$ and in the arbitrary plane of $\varphi = const.$, respectively. Note that the circular polarization means here the "interface" circular polarization defined in this plane. Thus, for oblique incidence, circular polarization commonly defined in the plane of the beam cross-section resolves into elliptic polarization in the interface plane and vice versa. For pure CR (CL) incident polarization $\widetilde{\chi}^{(i)\ -1}_{(R,L)} = 0$ and $\Delta_{CR} = 0$ ($\widetilde{\chi}^{(i)\ +1}_{(R,L)} = 0$ and $\Delta_{CL} = 0$), the CR (CL) coefficients are equal to C coefficients $t_{CR} = t_C$ and $r_{CR} = r_C$ ($t_{CL} = t_C$ and $r_{CL} = r_C$). At the same time the opposite CL (CR) coefficients $t_{CL}$ and $r_{CL}$ ($t_{CR}$ and $r_{CR}$) lead to the excitation of the CL (CR) components of the opposite polarization and with the exponential factors $\exp(+2i\varphi)$ ($\exp(-2i\varphi)$). All these coefficients are independent of $\varphi$ together with the sense of the circular polarization. Note that contrary to beam refraction, for beam reflection the "opposite" and "direct" component mean in fact the components of the same and opposite polarization as that of the incident beam, respectively.

For incidence other than normal and for arbitrary beam polarization both coefficients $t_C$ and $r_C$ are modified by the components $\Delta_{CR}$ and $\Delta_{CL}$ due to the XPC effect of the beams at the interface. That makes all coefficients $t_{CR}$, $t_{CL}$, $r_{CL}$ and $r_{CR}$ dependent on $\varphi$ through $\Delta_{CR}$ and $\Delta_{CL}$. For *critical incidence* of TIR $t_C = 1, r_C = 0$, $t_{CX} = -r_{CX} = -1$, although $\Delta_{CR} = \Delta_{CR}^{TIR}$ still in general differs from $\Delta_{CL} = \Delta_{CL}^{TIR}$:

$$\Delta_{CL}^{TIR} = -\widetilde{\chi}_{(R,L)}^{(i)} \exp(+2i\varphi) = \left(\Delta_{CR}^{TIR}\right)^{-1}. \tag{48}$$

If, however, the beam *incidence is normal*, then $\eta = 1$, $t_{CX} = 0 = r_{CX}$, $\Delta_{CR} = 0 = \Delta_{CL}$ so $t_{CR} = t_{CL} = t_p = t_s$ and $r_{CR} = r_{CL} = r_p = -r_s$ and the central ray of the beam, normal to the interface, is ruled only by geometrical optics.

All these characteristic features of beams of circular symmetry are explicitly shown in the diagonal/antidiagonal decomposition [12], alternative to the definitions (40)-(41):

$$\underline{t}_{(R,L)} = \underline{t}_C + \underline{t}_{CX} = t_C \begin{bmatrix} 1 & 0 \\ 0 & 1 \end{bmatrix} + t_{CX} \begin{bmatrix} 0 & \exp(-2i\varphi) \\ \exp(+2i\varphi) & 0 \end{bmatrix}, \tag{49}$$

$$\underline{r}_{(L,R)} = \underline{r}_C + \underline{r}_{CX} = r_C \begin{bmatrix} 1 & 0 \\ 0 & 1 \end{bmatrix} + r_{CX} \begin{bmatrix} 0 & \exp(-2i\varphi) \\ \exp(+2i\varphi) & 0 \end{bmatrix}. \tag{50}$$

For any polar $\vartheta^{(i)}$ and azimuthal $\varphi$ angle they resolves into the diagonal matrices $\underline{t}_C$ and $\underline{r}_C$ determining the "direct" beam component of the same polarization as that of the incident beam and into the XPC antidiagonal matrices $\underline{t}_{CX}$ and $\underline{r}_{CX}$ determining the opposite beam component. This is the reason why the coefficiens $t_{CX}$ and $r_{CX}$ may be called the XPC coefficients. Under beam transmission and reflection the related to them matrix elements $t_{CX} \exp(\pm 2i\varphi)$ and $r_{CX} \exp(\pm 2i\varphi)$ create, in the opposite beam components, additional vortices of the topological charge equal $\pm two$.

Positions of the excited vortices are determined by zeroes of the beam field and thus they depend on values of the XPC coefficients $t_{CX}$ and $r_{CX}$. For normal incidence $t_{CX} = r_{CX} = 0$ and the excited vortices are placed exactly in the centre of the coordinate system $k_X w_w = 0 = k_Y w_w$. For oblique incidence, however, the spectrum centre acquires additional phase shift $2^{-1/2} k^{(i)} \sin \theta^{(i)}$ and the exited vortex is displaced additionally by the opposite spectral shift $-2^{-1/2} k^{(i)} \sin \theta^{(i)}$ to the position where again $t_{CX} = -r_{CX} = 0$. For critical incidence, for example, the displacement of the excited vortex induces the change from the point where $t_{CX} = -r_{CX} = -1$ to the point where $t_{CX} = -r_{CX} = 0$.

The problem discussed here does not possess the symmetry of Cartezian coordinates in the plane $k_X - k_Y$ and, in general, no plane exists where the problem resolves into the standard Fresnel transmission or reflection and where the XPC coupling disappears. In spite of this the matrices (49)-(50) can be also rewritten, similarly to the decomposition (26)-(27) in the rectangular symmetry [17], into the zero-, first- and second- order terms with respect to $\varphi$:

$$\underset{=(R,L)}{t} = \begin{bmatrix} t_C & t_{CX} \\ t_{CX} & t_C \end{bmatrix} + it_{CX}\begin{bmatrix} 0 & -1 \\ 1 & 0 \end{bmatrix}\sin 2\varphi - 2t_{CX}\begin{bmatrix} 0 & 1 \\ 1 & 0 \end{bmatrix}\sin^2\varphi, \qquad (51)$$

$$\underset{=(L,R)}{r} = \begin{bmatrix} r_C & r_{CX} \\ r_{CX} & r_C \end{bmatrix} + ir_{CX}\begin{bmatrix} 0 & -1 \\ 1 & 0 \end{bmatrix}\sin 2\varphi - 2r_{CX}\begin{bmatrix} 0 & 1 \\ 1 & 0 \end{bmatrix}\sin^2\varphi. \qquad (52)$$

It is evident that the first g-o term in the CR/CL decomposition (51)-(52) takes over the role of the Fresnel matrices $\underset{=(p,s)}{t}$ and $\underset{=(p,s)}{r}$ in the TM/TE decomposition (26)-(27). Two next terms in Eqs (51)-(52) are just the XPC modifications to this g-o contribution. There are significant differences between the decomposition (26)-(27) in the TM/TE basis and the decomposition (51)-(52) in the CR/CL basis. The first-order terms show the additional phase shift by $\pi/2$ and the second-order (antidiagonal) term is caused by the XPC effect, contrary to the case of the TM/TE case. Moreover, the zero-order (g-o) matrices in the decomposition (51)-(52) are not diagonal. That means that the XPC effects directly enter into the g-o beam field specification and excite the nonzero field distribution in the opposite beam component. However, this XPC contribution to the g-o beam field still does not change the winding number of the beams and disregards the vortex excitation at the interface. This process is out of reach by g-o approach to the problem under consideration.

As the g-o terms do not depend on $\varphi$, the g-o transmission and reflection usually dominates over the weaker first-order and second-order terms for small values of $\varphi$. It is only one exception – normal incidence of the beams. In this case $t_{CX} = 0 = r_{CX}$ and the g-o matrices in Eqs (51)-(52) become diagonal and independent of the XPC effect. That implies that, for normal incidence, optical vortices are undistorted and clearly visible in both, spectral and configuration, domains. Still from Eqs (49)-(52) it is evident that, for any beam incidence, the first-order and second-order XPC terms - not the g-o terms - in the decomposition (51)-(52) - describe the process of creation of optical vortices at the interface. But this aspect of the beam-interface interactions is here interesting the most from the fundamental and application points of view.

## 6. LAGUERRE-GAUSSIAN MODES OF THE INTERFACE

Potential of the relations (49)-(52) can be checked by the analysis of the ELG/PELG beams at the interface. They are defined by appropriate differentiation (9) of the fundamental Gaussian beam field in the configuration domain. In the spectral domain, that definition is equivalent to the algebraic expression:

$$\widetilde{G}_{p,l}^{(EL)}(\kappa,\overline{\kappa},Z') = (iw_w)^{2p+l}\kappa^{p+l}\overline{\kappa}^p \widetilde{G}_{0,0}^{(EL)}(\kappa,\overline{\kappa},Z'), \tag{53}$$

where $\kappa = 2^{-1/2}(k_X + ik_Y) = \kappa_\perp \exp(i\varphi)$, $\overline{\kappa}$ means complex conjugate of $\kappa$, $\kappa_\perp^2 = \kappa\overline{\kappa}$, $\widetilde{G}_{p,l}^{(EL)}$ and $\widetilde{G}_{0,0}^{(EL)} = 2\pi \exp(-\kappa_\perp^2 v^2)$ are the Fourier transformed ELG/PELG and Gaussian beams, respectively. In the polar coordinates $\kappa_\perp$ and $\varphi$ the above definition reads:

$$\widetilde{G}_{p,l}^{(EL)}(\kappa,\overline{\kappa},Z') = (i\kappa_\perp w_w)^{2p+l}\widetilde{G}_{0,0}^{(EL)}(\kappa,\overline{\kappa},Z')\exp(il\varphi), \tag{54}$$

where $\varphi = (1/2)\ln(\kappa/\overline{\kappa})$ and the identity $\kappa^{p+l}\overline{\kappa}^p = \kappa_\perp^{2p+l}\exp(il\varphi)$ was applied [12].

For incidence of the beam field $\widetilde{\underline{E}}^{(i)} = [a_R, a_L]^T \widetilde{G}_{p,l}^{(EL)}$ of the mode $\widetilde{G}_{p,l}^{(EL)}$ and of arbitrary polarization $\widetilde{\chi}^{(i)}(R,L) = \widetilde{a}_R/\widetilde{a}_L$, specified by the in general nonuniform and complex beam components $\widetilde{a}_R$ and $\widetilde{a}_R$, the relations (51)-(54) yield [12]:

$$\begin{bmatrix}\widetilde{E}_R^{(t)}\\ \widetilde{E}_L^{(t)}\end{bmatrix} = t_C \begin{bmatrix}\widetilde{a}_R\\ \widetilde{a}_L\end{bmatrix}\widetilde{G}_{p,l}^{(EL)} + t_{CX}\begin{bmatrix}\widetilde{a}_L \widetilde{G}_{p+1,l-2}^{(EL)}\\ \widetilde{a}_R \widetilde{G}_{p-1,l+2}^{(EL)}\end{bmatrix}, \tag{55}$$

$$\begin{bmatrix}\widetilde{E}_L^{(r)}\\ \widetilde{E}_R^{(r)}\end{bmatrix} = r_C \begin{bmatrix}\widetilde{a}_R\\ \widetilde{a}_L\end{bmatrix}\widetilde{G}_{p,l}^{(EL)} + r_{CX}\begin{bmatrix}\widetilde{a}_L \widetilde{G}_{p+1,l-2}^{(EL)}\\ \widetilde{a}_R \widetilde{G}_{p-1,l+2}^{(EL)}\end{bmatrix}. \tag{56}$$

If the incident ELG beam has only *one* component of, say, CR (CL) circular polarization, the action of the interface yields also the ELG beam but with *two* orthogonal components. The first component has the same CR (CL) polarization as the incident beam is excited by the g-o terms in Eqs (49)-(50) and the second component, with the opposite CL (CR) polarization, is excited by the XPC terms in these equations. Contrary to the first component, which preserves indexes of the incident beam ($p \to p$ and $l \to l$), in the second, opposite component the radial index is changed by one ($p \to p \mp 1$) and the azimuthal index is changed by two ($l \to l \pm 2$). So, within the definitions assumed for the coefficient $t_{CX}$ and the ELG beam field $\widetilde{G}_{p,l}^{(EL)}$, the order $N^{(EL)} = 2p+l$ of all ELG beam components remains unchanged ($N^{(EL)} \to N^{(EL)}$).

The problem appears different for oblique incidence, as the beam transmission/reflection for oblique incidence is affected by the narrow azimuthal range of the beam spectrum. Then the relations (51)-(52) are more appropriate in this case. A position of the vortex excited by the XPC effect is shifted in this case by $-2^{-1/2}k^{(i)}\sin\theta^{(i)}$ to the point $(k_X w_w, k_Y w_w) = (0,0)$. Moreover, the excited vortex could be visible only when the beam is sufficiently narrow to cover this point by its spectrum range. Still the relations of the global topological charge of the beams remains the same as for normal incidence. Note also that the spectral amplitudes of both field components are further modified by the (direct) $t_C$, $r_C$ and (cross) $t_{CX}$, $r_{CX}$ coefficients, respectively, as both of them are dependent on $\vartheta^{(i)}$. Therefore the change of radial indices $p$ depends on the defined form (53)-(54) of the modes $\widetilde{G}_{p,l}^{(EL)}$ and on the definitions (20)-(21) of the coefficients $t_{CX}$ and $r_{CX}$ assumed in Eqs (55)-(56).

In contrast to the EHG/PEHG beam decomposition (30)-(33), the ELG/PELG decomposition (55)-(56) is exact. All terms, including the first-order and second-order terms, are accounted for in this decomposition. It is also highly symmetric. In spite of the obvious difference between C and CX coefficients the relations (55)-(56) differentiate between CR and CL polarization of the incident beam only by the changes of the indices $p$ and $l$ in the XPC term.

It is well known that the LG beams carry the total angular momentum (TAM) composed of the spin angular momentum (SAM) and the orbital angular momentum (OAM) [23-25]. For CR or CL beam polarization the beam helicity $\sigma$ equals $+1$ or $-1$, respectively, and SAM of the beam equals $\sigma\hbar$ per photon. Similarly, for LG beams of the winding number $l$ OAM of the beam equals $l\hbar$ per photon. Note that TAM, SAM and OAM are measured with respect to a normal to the interface. Thus, for the incident ELG/PELG beam with CR/CL polarization and of the winding number $l$, the $Z$ component of TAM per photon equals $(\sigma+l)\hbar$; $(1+l)\hbar$ for the beam of CR polarization and $(-1+l)\hbar$ for the beam of CL polarization.

Reasoning per analogy, in the relations (55)-(56) the $Z$ component of TAM per photon equals the same amount $(\sigma+l)\hbar$ independently for the transmitted and reflected ELG/PELG beams as well as for any -CR or CL - component of these beams. Therefore the $Z$ component of TAM per photon of the ELG/PELG beam with the CR/CL polarization is conserved during beam-interface interactions for each photon of the beam. Moreover, taking into account that the total number of photons survives during the beam-interface linear interactions, that also means that the global TAM integrated over the ELG/PELG beam cross-sections in the interface plane with CR/CL polarization seems to be also conserved.

The case of the normally incident CL polarized $ELG_{2,4}$ beam collimated at the interface ($z=0=Z_w$) is shown in figure 4. The field phase distributions in the configuration and spectral domains are shown for the incident beam and for two orthogonal components of the reflected beam. All the phase distributions are circularly symmetric in the interface plane with the vortex singularity placed exactly in the centre of the figure. The beam phase change along a closed loop about this point equal to $4\times 2\pi$ for the incident beam and for the CR component of the reflected beam. However it equals to $2\times 2\pi$ for the CL component of the reflected beam, as exactly predicted by Eq. (56). Excitation efficiency of this opposite beam component amounts about one order in beam intensity less than that for the direct component of the (CR) polarization. Results of numerical simulations shown in figure 4 for beam reflection are entirely consistent with those for beam transmission given in [12].

The case of critical incidence of the CL polarized $ELG_{2,4}$ beam collimated at the interface is shown in figure 5 through the beam field intensity and phase distribution in the spectral domain. As for the PEHG beams the symmetry specific to normal incidence is broken by (i) the field projection on the interface plane, (ii) the asymmetry of the Fresnel coefficients with respect to $\vartheta^{(i)}$ and by (iii) the narrow azimuthal range of the incident beam spectrum. In the spectral domain, for the reason (i) even the incident beam intensity shows deviations from the elliptic symmetry (cf. figure 5a). For the reason (ii) the reflected beam spectra in both polarization components are almost annihilated for rays of $\vartheta^{(i)}$ less that the critical angle $\theta_c^{(i)}$ (cf. figure 5b and figure 5c). Moreover, for the reason (iii) the first-order and the second-order contributions to the reflected field are still dominated by the first, g-o terms in the field decomposition (55)-(56), in spite of the fact that $r_C = 0$ exactly at $\vartheta^{(i)} = \theta_c^{(i)}$. These effects lead to severe beam deformations in the configuration domain. The narrow beam (here its radius is of the order of one wavelength) is so badly deformed in the configuration domain that any concise interpretation of what it really shows would be rather difficult.

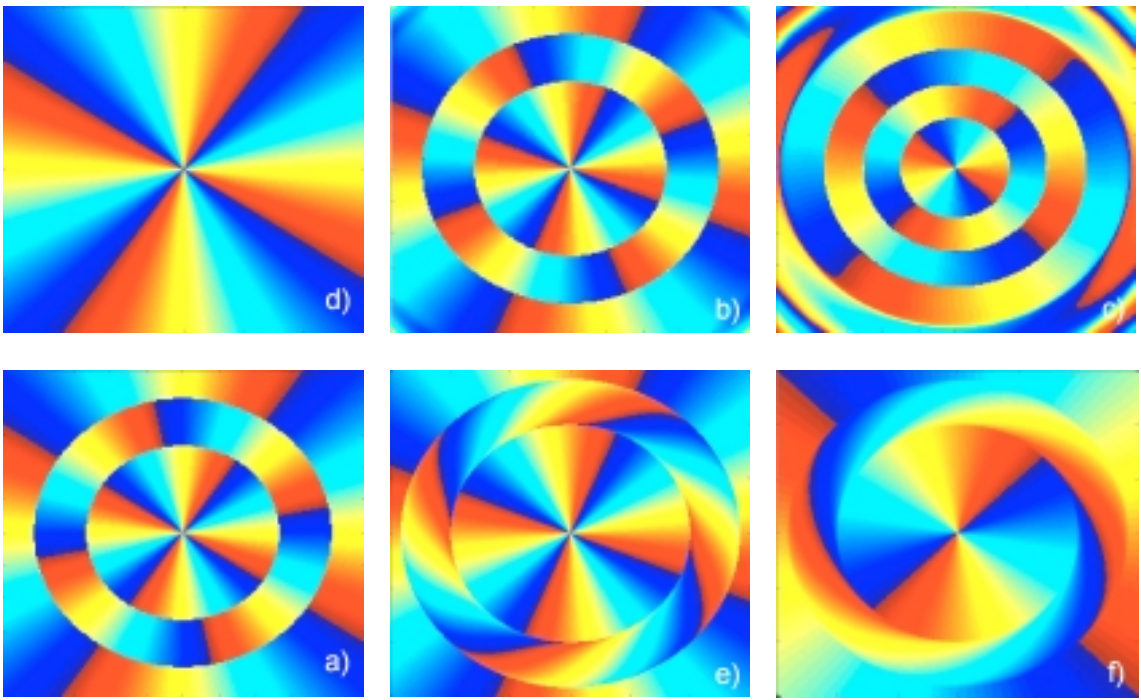

**Figure 4.** Phase transverse distribution of the ELG beam components in the configuration (a)-(c) and spectral (d)-(f) domains evaluated in the plane of the interface for $z = 0$. The case of normal incidence ($\theta^{(i)} = 0^o$) of the beam collimated at the interface is displayed for: (a) the incident beam of the $ELG_{2,4}$ pattern and of CL polarization, (b) the reflected beam CR component of the $ELG_{2,4}$ pattern, (c) the reflected beam CL component of the $ELG_{3,2}$ pattern. Counterparts of the figures (a)-(c) in the spectral domain are depicted in the figures (d)-(f). The ELG patterns in the figures (b), (e) and (c), (f) are modified by the reflection coefficients $r_C$ and $r_{CX}$, respectively.

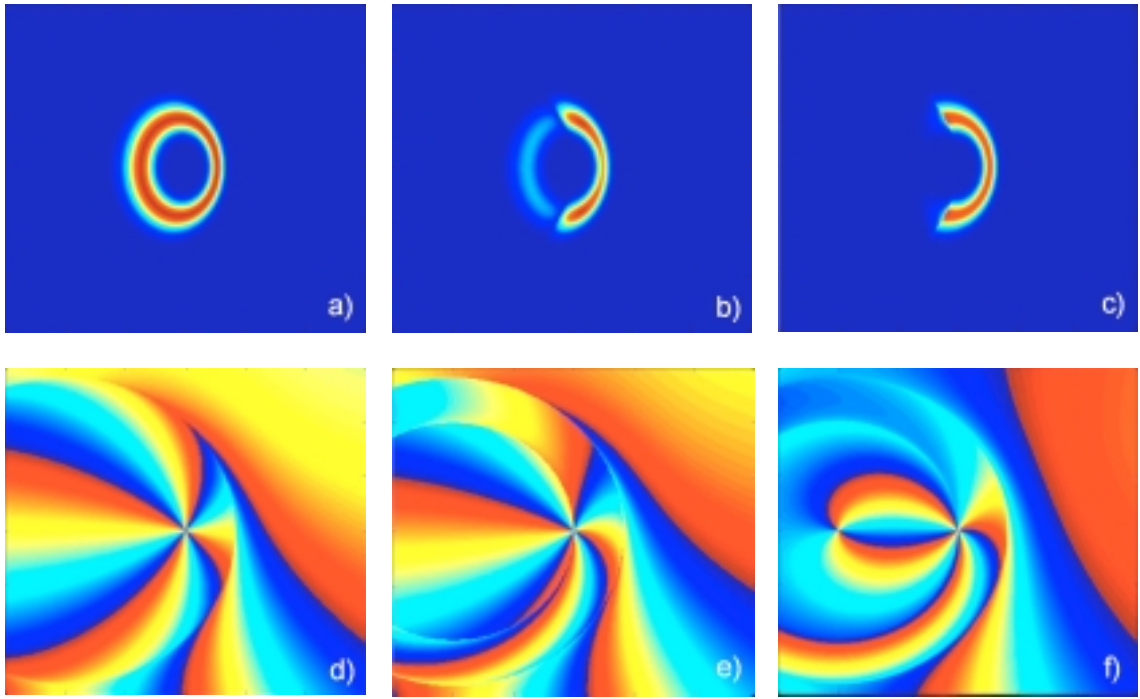

**Figure 5.** Intensity (a)-(c) and phase (d)-(f) transverse distribution of the ELG beam spectral components at the interface for $z = 0$. The case of critical incidence ($\theta_c^{(i)} = 45^o$) of the beam collimated at the interface is displayed for: intensity (a) and phase (d) of the incident beam of the PELG$_{2,4}$ pattern and of CL polarization, intensity (b) and phase (e) of the reflected beam CR component of the PELG$_{2,4}$ pattern, intenisty (c) and phase (f) of the reflected beam CL component of the PELG$_{3,2}$ pattern. The ELG patterns are modified by the coefficients $r_C$ and $r_{CX}$.

However, even in this case, the picture of transverse field distribution is still quite clear in the spectral domain because the symmetry of the beam phase structure, known from the case of normal incidence, survives here to a large extent. The on-axis vortex of the topological charge equal to four is placed exactly in the central ray of the beam (for $\vartheta = \theta_c^{(i)}$) for the incident beam (cf. figure 5d), as well as for the reflected beam component of the direct CR polarization (cf. figure 5e). Moreover, the vortex of same type can be spotted in the spectrum of the opposite (CL) component of the reflected beam (cf. figure 5f). Still the global topological charge of the vortices excited by the XPC effect equals to two as expected, because the second, off-axis vortex of the opposite charge equal to minus two is also excited (cf. figure 5f). A position of this additional vortex is shifted in the incidence plane by $-k^{(i)} w_w \sin\theta^{(i)}$ to the point $k_X w_w = 0$, that is where $r_{CX} = 0$ and the spectral intensity is low. Therefore a trace of this vortex in the configuration domain is weak and difficult to observe. Nevertheless, it is the first, to the author's knowledge, explicit analytic evaluation and its numerical verification of the off-axis position of a vortex excited at the interface.

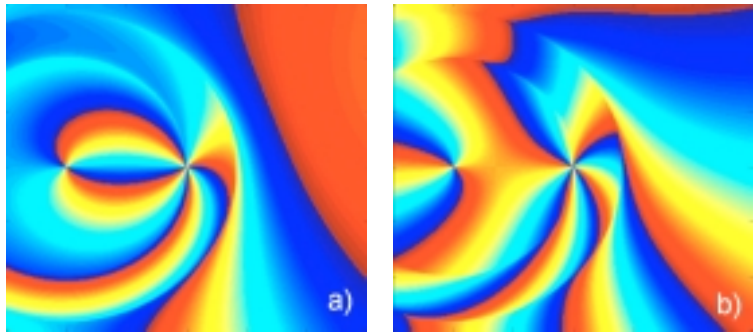

**Figure 6.** Optical vortex excitation by the XPC effect for the incident beam of the $ELG_{2,4}$ shape. Phase distribution for the collimated beam (a) of *CL polarization* of the incident beam and the reflected beam component; the total topological charge equals $4-2$, (b) of *CR polarization* of the incident beam and the reflected beam component; the total topological charge equals $4+2$. In both cases the topological charge of the incident beam equals $4$. All other data are the same as for figure 5.

Numerical results presented in figure 5 not only confirm predictions of Eqs (55)-(56) but also directly explain the mechanism of vortex creation or annihilation caused by the XPC effect at the interface. The change by two of the topological charge in the opposite beam component, as predicted by Eqs (55)-(56), results from creation of the additional, off-axis in general, vortex of the topological charge equal to plus or minus two. The topological charge of this additional vortex adds to or subtracts from the topological charge of the vortex which mimics that of the incident beam. The process depends directly on the relative sign of the phase rotation, indicated by the signs of the azimuthal index $l$, with respect to the sign of rotation of the field vector, distinguished by the beam CR or CL polarization (cf. figure 6). Figure 7 displays helical rotations of the phase structure of the propagating reflected beam about its vortex singularities. During beam propagation the beam phase structure rotates about the phase singularities of their precisely robust positions.

Figures 5-7 visualise the main numerical result of the paper. The process of vortex excitation at the interface is presented in all its spectral aspects. The figures show explicitly its dependence on beam incidence and indicate positions of the excited vortices precisely at points obtained entirely by the theoretical analysis presented. The process depends on the range of beam spectrum and become efficient when this range covers points of vortex centre positions. It always happens for normal incidence. For oblique incidence it depends on a balance between the beam incidence and the beam width – when the incidence angle becomes larger, the beam

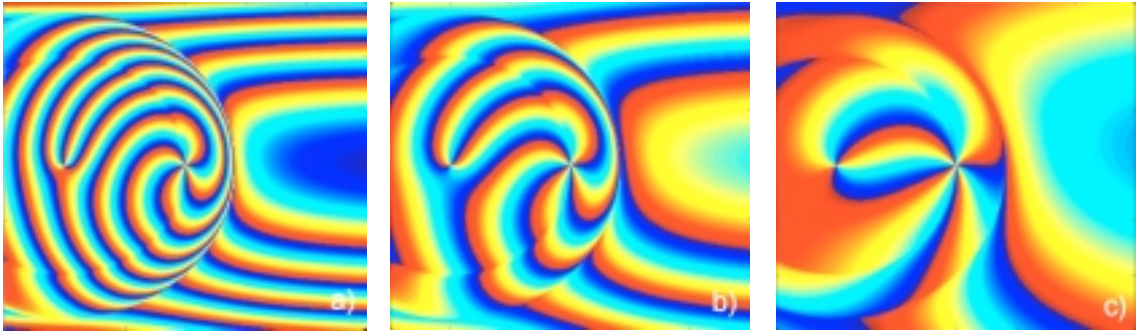

**Figure 7.** Vortex phase evolution during propagation of the reflected CL polarized beam of the $ELG_{3,2}$ shape shown in the spectral domain for its waist placed at: $z/z_D =$: (a) $-1.0$, (b) $-0.5$, (c) $-0.125$. The case of $z/z_D = 0.0$ is shown in figure 5c and 6a. All other data are the same as for figure 5.

width should become smaller. It seems that the process of vortex excitation under oblique incidence is a characteristic feature of behaviour of nonparaxial beams at the interface.

Note finally that, as in the case of incidence of EHG/PEHG beams of TM or TE polarization, for incidence of ELG/PELG beams of CR or CL polarization the transmitted/reflected beam is composed also of the ELG/PELG modes in the two opposite polarization components. In spite of the beam deformations the beam indices are exactly specified without any approximation to the transmission and reflection matrices (49)-(52). Therefore the ELG/PELG beams may be regarded as *exact* normal modes of the dielectric interface - *exact* because their relations (55)-(56) at the interface are exact.

## 7. ADDITIONAL COMMENTS

The analysis applied in this paper for the beam-interface interactions is quite general. In Refs [12], [16] and [17] the XPC effects were numerically analysed mainly for seemingly the most interesting cases of normal and critical incidences at the interface. However, they certainly exist in other configurations and incidence angles. By use of appropriately defined transfer and scattering matrices the method can be directly generalised into cases of beams at any planar multilayered structure [26]. That includes structures composed of nonlinear media as well, where the XPC effect is augmented by effects of self- and cross-phase modulation [21]. The XPC effects were also recently reported for a linearly polarized fundamental Gaussian beam incident upon the interface at a Brewster angle [27].

One issue still left to be commented is the interpretation of substantial deformations of the PELG beams in the configuration domain. It seems, at least in principle, that it is possible to evaluate the PELG beam deformations by analogy to the prescription of the beam shifts given in sections 3 and 4 for the PEHG beams. The g-o coefficients $r_C$ and $r_{CX}$ still depend on $\vartheta^{(i)}$ and their modification terms $\Delta_{CR}$ and $\Delta_{CL}$ still depend on $\varphi$. However, the interpretation of these effects is not so straightforward as the rectangular symmetry necessary in this case is blurred by the cylindrical symmetry of the obliquely incident PELG beams.

On the other hand, it is still possible to analyse the deformations of the PELG beams directly by use of the effects or transformations of nonspecular transmission and reflection of the PEHG beams. These effects can be evaluated in rectangular coordinate system accordingly to the procedure suggested in [15-17] and further separated into two transformations related to the beam amplitude and to the beam polarization [17]. While the polarization transformations describe the action of the interface, or in general of any multilayered structure, in terms of Lorentz transformations, the amplitude transformations yield spatial beam shaping [17].

As both families of these beams form complete biorthogonal sets of solutions of the paraxial wave equation, any PELG beam can be expressed by a linear combination of the PEHG beams and vice versa [23, 24]. Moreover, the expansion coefficients of such decompositions for the elegant HG and LG beams are determined by their counterparts for standard beams [3, 28], a feature being characteristic even for more general classes of beams [29, 30]. More specifically, any PELG beam of the order N=2p+l and dependent on in $\varsigma$ and $\bar{\varsigma}$ can be also decomposed into the set of PEHG beams of the same order $N = m + n$, with the expansion coefficients $b(m,n,k)$ defined by the similar decomposition of diagonal PEHG beams dependent on $X' = 2^{-1/2}(X+Y)$ and $Y' = 2^{-1/2}(X-Y)$:

$$G_{p,l}^{(EL)}(\varsigma,\bar{\varsigma},Z') = \sum_{k=0}^{N} i^k b(m,n,k) G_{N-k,k}^{(EH)}(X,Y,Z'), \qquad (57)$$

$$G_{m,n}^{(EH)}(X',Y',Z') = \sum_{k=0}^{N} b(m,n,k) G_{N-k,k}^{(EH)}(X,Y,Z'), \qquad (58)$$

where $m = p$, $n = l + p$ for $m < n$ and $m = l + p$, $n = p$ for $m > n$ [23, 24]. To estimate the PELG beam displacements or deformations, one may analyse these effects first for PEHG beams along the lines described in [15-17]. Next, using the expansion coefficients determined by Eq. (58), one could obtain the deformed rather than displaced PELG beams by incorporating these effects into the mode components of the expansion (57).

For beams sufficiently wide to be within a deep paraxial range and for fundamental Gaussian incidence, direct application of the method developed in [15-17] is adequate in analysing the effects of beam nonspecular transmission and reflection at the interface. The method covers all cases ranged from internal reflection to partial transmission, fully accounts for vectorial and 3D nature of beam fields and yields analytical expressions for beam polarization, intensity and phase. Moreover, it relates the nonspecular modifications of beam mode shapes with beam polarization through the polarization parameter (22) or the XPC effect [12], which accounts for all cases of beam polarization, including nonuniform polarization as well. The method provides, in the spectral and configuration domains, the description of all the first-order and second-order nonspecular effects, including those affecting the beam centre position, beam propagation direction and width, as well as the on-axis amplitude and phase [15-17].

## 8. SUMMARY AND CONCLUSIONS

Beam interaction with the dielectric interface has been analysed in the rectangular and cylindrical coordinate systems. It was shown that the Fresnel transmission/reflection coefficients should be replaced by their generalisations appropriate for 3D beams. All, the g-o or zero-order, first-order and second-order, contributions to the beam field are governed by these new coefficients - $t_C$, $r_C$ of the g-o type and $t_{CX}$, $r_{CX}$ of the XPC type. Accordingly to the definitions (20)-(21) and (42)-(43) they are composed of sums and differences of the common p and s Fresnel coefficients and satisfy only two scalar field continuity relations at the interface: $t_C = 1 - r_C$ and $t_{CX} = -r_{CX}$, in opposition to four scalar relations necessary for the Fresnel coefficients. As far as a single central ray of the beam is considered, $t_{CX} = 0 = r_{CX}$ and the XPC effects disappear for normal incidence. On the other hand $t_{CX} = -1 = -r_{CX}$ for critical incidence and the scattering part of the g-o effects disappears in this case.

In the g-o beam field the XPC effects are absent in the rectangular coordinates and present in the cylindrical coordinates. Still, even in the latter case, optical vortices are not excited in the g-o beam field. The XPC effects are incorporated effectively in the first-order and second-order contributions to the transmission and reflection matrices. It was shown that the action of these effects leads to creation of higher-order modes in the opposite - to the polarization state of the incident beam - field component and can be explicitly described by changes of mode indices. Meanwhile in the rectangular coordinates these changes can be analytically proved only within the first-order approximation to the beam field, the same analysis carried out in the cylindrical coordinates appears analytically exact. The analysis presented is quite general and valid for arbitrary, linear and elliptic, uniform and nonuniform, polarization of the incident beam.

Two sets of elegant higher-order beams have been considered as beam normal modes at the interface: EHG beams of linear polarization and ELG beams with circular polarization. As the interface couples polarization and spatial (in amplitude and phase) structures of the beams, normal modes have been treated as scalar modes dressed by linear polarization for EHG modes and circular polarization for ELG modes. Both of them have been analysed in their normal incidence; in oblique incidence their PEHG and PELG projections on the interface plane have been considered instead. Under normal incidence the EHG beams acquire the change by one in both their indices. However for oblique incidence one from these two index changes - that in the plane of incidence - is gradually replaced by the net longitudinal spatial shifts of the whole PEHG beam field structure.

The case of ELG/PELG beams appeared especially interesting as the analysis revealed creation of additional optical vortices at the interface. Their spectral positions, placed in the incidence plane $k_Y w_w = 0$, are spectrally displaced from the incident vortex placement $k_X w_w = k w_w \sin\theta^{(i)}$ predicted by g-o optics to the position $k_X w_w = 0$, where $t_{CX} = -r_{CX} = 0$. The process of vortex excitation is efficient when the range of beam spectra covers the spectral positions of the excited vortices. Note that exactly the same shift of spectral position of the whole beam phase structure is also observed for the PEHG beams. It was also shown in the figures that, for the normal and oblique incidence, the changes of SAM related to the beam component of opposite polarization are compensated by the changes of OAM related to the excited vortices immersed in the beam field structures. The process leaves the normal to the interface TAM of the beam unchanged.

The method applied in the numerical simulations is based on direct integration of Maxwell equations. Incidence of narrow beams with their cross-section radii of the order of one wavelength was considered. The simulations entirely confirmed analytical predictions. Only the case of beam reflection was numerically analysed. Numerical treatment of beam transmission, that for a change based directly on the equations derived and complementary to what is already available in [12], will be reported elsewhere [31]. The analysis has been presented for the case of a single interface but its generalisation to the case of any planar multilayer is straightforward.

**REFERENCES**

[1] M. Born and E. Wolf, *Principles of Optics*, 7th ed. (Cambridge University Press, Cambridge, England, 1999).

[2] P. Yeh, *Optical Waves in Layered Media* (Wiley, New York, 1976).

[3] W. Nasalski, *Optical Beams at Dielectric Interfaces – fundamentals* (IPPT PAN, Warsaw, 2007).

[4] J. Picht, "Beitrag zur Theorie der Totalreflexion", Ann. Phys. Leipzig **5**, 433-496 (1929).

[5] F. Goos and H. Hänchen, "Ein neuer and fundamentaler Versuch zur Totalreflexion", Ann. Phys. (Leipzig) **1**, 333-345 (1947).

[6] F. I. Fedorov, "K teorii polnovo otrazenija". Dokl. Akad. Nauk SSSR **105**, 465-467 (1955).

[7] C. Imbert, "Calculation and experimental proof of the transverse shift induced by total internal reflection of a circularly polarized light beam", Phys. Rev. D **5**, 787-796 (1972).

[8] H. Okuda and H. Sasada, "Significant deformations and propagation of Laguerre-Gaussian beams reflected and transmitted at a dielectric interface", J. Opt. Soc. Am. A **25**, 881-890 (2008).

[9]   R. Zambrini and S. M. Barnett, "Quasi-intrinsic angular momentum and the measurement of its spectrum", Phys. Rev. Lett. **96**, 113901 (2006).

[10]  G. Molina-Terriza, J. P. Torres, and L. Torner, "Management of the angular momentum of light: preparation of photons in multidimensional vector states of angular momentum", Phys. Rev. Lett. **88**, 013601 (2002).

[11]  L. Torner, J. P. Torres, S. Carrasco, "Digital spiral imaging", Opt. Express **13**, 873-881 (2005).

[12]  W. Nasalski, „Polarization versus spatial characteristics of optical beams at a planar isotropic interface", Phys. Rev. E **74**, 056613 (2006).

[13]  A. E. Siegman, "Hermite-Gaussian functions of complex argument as optical beam eigenfunctions", J. Opt. Soc. Am. **63**, 1093-1995 (1973).

[14]  A. E. Siegman, *Lasers* (University Science Books, Mill Valley, CA, 1986).

[15]  W. Nasalski, "Longitudinal and transverse effects of nonspecular reflection", J. Opt. Soc. Am. A **13**, 172-181 (1996).

[16]  W. Nasalski, "Three-dimensional beam reflection at dielectric interfaces", Opt. Commun. **197**, 217-233 (2001).

[17]  W. Nasalski, "Amplitude-polarization representation of three-dimensional beams at a dielectric interface", J. Opt. A: Pure Appl. Opt. **5**, 128-136 (2003).

[18]  S. Saghafi and C. J. R. Sheppard, "Near field and far field of elegant Hermite-Gaussian and Laguerre-Gaussian modes", J. Mod. Opt. **45**, 1999-2009 (1998).

[19]  J. Enderlein and F. Pampaloni, "Unified operator approach for deriving Hermite-Gaussian and Laguerre-Gaussian laser modes", J. Opt. Soc. Am. A **21**, 1553-1558 (2004).

[20]  F. I. Baida, D. Van Labeke, J-M. Vigoureux, "Numerical study of the displacement of a three-dimensional Gaussian beam transmitted at total internal reflection. Near-field applications", J. Opt. Soc. Am. A **17**, 858-865 (2000).

[21]  W. Nasalski, "Modelling of beam reflection at a nonlinear-linear interface", J. Opt. A: Pure Appl. Opt. **2**, 433-441 (2000).

[22]  A. Aiello and J. P. Woerdman, "Role of beam propagation in Goos-Hänchen and Imbert-Fedorov shifts", Opt. Lett. **33**, 1437-1439 (2008).

[23]  L. Allen, M. W. Beijersbergen, R. J. C. Spreeuw, and J. P. Woerdman, „Orbital angular momentum of light and the transformation of Laguerre-Gaussian laser modes", Phys. Rev. A **45**, 8185-8189 (1992).

[24]  L. Allen, M. J. Padgett and M. Babiker, "The orbital angular momentum of light", Progress in Optics **39**, 291-372 (1999).

[25]  A. T. O'Neil, I. Mac Vicar, L. Allen, and M. J. Padgett, "Intrinsic and extrinsic nature of the orbital angular momentum of a light beam", Phys. Rev. Lett. **88**, 053601 (2002).

[26]  W. Nasalski, "Three-dimensional beam scattering at multilayers: formulation of the problem", J. Tech. Phys. **45**, 121-139 (2004).

[27] A. Köházi-Kis, "Cross-polarization effects of light beams at interfaces of isotropic media", Opt. Commun. **253**, 28-37 (2005).

[28] Y. Pagani and W. Nasalski, "Diagonal relations between elegant Hermite-Gaussian and Laguerre-Gaussian beam fields", Opto-Electron. Rev. **13**, 53-62 (2005).

[29] E. G. Abramochkin and V. G. Volostnikov, "Generalized Gaussian beams", J. Opt. A: Pure Appl. Opt. **6**, S157-S161 (2004).

[30] M. A. Bandres, "Elegant Ince-Gaussian beams", Opt. Lett. **29**, 1724-1726 (2004).

[31] W. Szabelak and W. Nasalski, "Excitation of optical vortices at a dielectric interface versus a beam incidence angle", Proc. SPIE 7141, 714115 (2008).